# Photon retention in coherently excited nitrogen ions

Jinping Yao[1,#], Luojia Wang[2,#], Jinming Chen[1,3,4], Yuexin Wan[1,3], Zhihao Zhang[1,3,4], Fangbo Zhang[1,3], Lingling Qiao[1], Shupeng Yu[1,3], Botao Fu[1,3,4], Zengxiu Zhao[5], Chengyin Wu[6], Vladislav V. Yakovlev[7], Luqi Yuan[2,*], Xianfeng Chen[2,8,9,10,*], and Ya Cheng[1,10,11,*]

[1]*State Key Laboratory of High Field Laser Physics and CAS Center for Excellence in Ultra-intense Laser Science, Shanghai Institute of Optics and Fine Mechanics (SIOM), Chinese Academy of Sciences (CAS), Shanghai 201800, China*

[2]*State Key Laboratory of Advanced Optical Communication Systems and Networks, School of Physics and Astronomy, Shanghai Jiao Tong University, Shanghai 200240, China*

[3]*University of Chinese Academy of Sciences, Beijing 100049, China*

[4]*School of Physical Science and Technology, ShanghaiTech University, Shanghai 200031, China*

[5]*Department of Physics, National University of Defense Technology, Changsha 410073, China*

[6]*State Key Laboratory for Mesoscopic Physics, School of Physics, Peking University, Beijing 100871, China*

[7]*Texas A&M University, College Station, TX 77843, USA*

[8]*Shanghai Research Center for Quantum Sciences, Shanghai 201315, China*

[9]*Jinan Institute of Quantum Technology, Jinan 250101, China*

[10]*Collaborative Innovation Center of Light Manipulations and Applications, Shandong Normal University, Jinan 250358, China*

[11]*Collaborative Innovation Center of Extreme Optics, Shanxi University, Taiyuan, Shanxi 030006, China*

[#]These authors contributed equally to this work.

[*]Corresponding authors: yuanluqi@sjtu.edu.cn; xfchen@sjtu.edu.cn; ya.cheng@siom.ac.cn

**Abstract:**

Quantum coherence in quantum optics is an essential part of optical information processing and light manipulation. Alkali metal vapors, despite the numerous shortcomings, are traditionally used in quantum optics as a working medium due to convenient near-infrared excitation, strong dipole transitions and long-lived coherence. Here, we proposed and experimentally demonstrated photon retention and subsequent re-emittance with the quantum coherence in a system of coherently excited molecular nitrogen ions ($N_2^+$) which are produced using a strong 800 nm femtosecond laser pulse. Such photon retention, facilitated by quantum coherence, keeps releasing directly-unmeasurable coherent photons for tens of picoseconds, but is able to be read-out by a time-delayed femtosecond pulse centered at 1580 nm via two-photon resonant absorption, resulting in a strong radiation at 329.3 nm. We reveal a pivotal role of the excited-state population to transmit such extremely weak re-emitted photons in this system. This new finding unveils the nature of the coherent quantum control in $N_2^+$ for the potential platform for optical information storage in the remote atmosphere, and facilitates further exploration of fundamental interactions in the quantum optical platform with strong-field ionized molecules.

Quantum coherence plays a central role in modern developments of quantum optics [1-3], which is of fundamental importance for understanding and predicting many initially counter-intuitive physical phenomena such as electromagnetically induced transparency [4-7], lasing without inversion [8-10], and mirrorless laser-like emission [11-13]. In particular, the light storage with quantum coherence, or the optical quantum memory, is of essential importance for achieving optical information storage in quantum communications [14-20]. Yet, many existing proposals in studying quantum coherence require alkali metal vapor medium, which possibly hinders the broader range of applications.

For several decades, alkali metal atoms have been serving as a backbone of quantum optical platforms [21-23] due to a number of unique properties such as sufficiently long coherence time, the lack of inter-vibrational relaxation, and easy access to laser wavelengths capable of the resonant excitation. Alkali metal dimers, which can be created through heating atoms at a high temperature, are another promising candidate for coherent quantum control [24-27]. Nevertheless, it would be of great importance to explore alternative solutions involving more abundant elements and molecules which are also less chemically reactant and safe to use. In particular, molecular nitrogen ions ($N_2^+$), which can easily be produced from atmospheric nitrogen using high-intensity femtosecond pulses, emerge as the intriguing model system for exploring quantum coherence effects from abundant energy levels that exhibit rich electron and molecular dynamics [28-34]. In such a system, the energy diagram is significantly different from that of an atom, i.e., energy differences between lower electronic states of molecular ions are located in near-ultraviolet, visible and near-infrared spectral regions, making it easy to efficiently manipulate the quantum coherence between multiple states by using commonly available lasers. Electronic, vibrational, and rotational coherences are created in molecular ions when electron instantaneously escapes from nucleus in a strong laser field [29]. However, the quantum coherence in strong-field-ionized molecular ions remains largely unexplored, opening the venue for achieving optical information storage and for exploring coherent quantum control in excited nitrogen ions.

In this paper, we unveil the nature of photon retention in $N_2^+$ both theoretically and experimentally. The proposed idea of achieving the photon retention using $N_2^+$ quantum optical platform is briefly illustrated in Fig. 1. Three electronic states of $N_2^+$, i.e., $X^2\Sigma_g^+$, $A^2\Pi_u$ and $B^2\Sigma_u^+$ (abbreviated as *X*, *A* and *B*), are considered to construct a three-level system as shown in Fig. 1(b). An intense laser at 800 nm is used to ionize molecular $N_2$ into $N_2^+$ and to pump $N_2^+$ into the excited *A* state. The same laser also generates the quantum coherence between *X* and *A* states, which keeps re-emitting photons near 800 nm for the timescale of picoseconds. The re-emitting is extremely weak that it is hardly measured directly. Yet, once another laser at 1580 nm is injected after the picosecond delay, we find that it can readout such re-emitted photons by promoting the two-photon resonance, which, in its turn, leads to an experimentally observed strong emission at 329.3 nm. In the same time, the population on *B*(v=4) level significantly amplifies the contribution of the coherently re-emitted photons, resulting in strong 329.3 nm emission, which persists even tens of picosecond after the initial excitation by the 800 nm laser. This report exhibits important coherent quantum control of photons in $N_2^+$, i.e., photon retention, which is fundamentally different from quantum memory for long lifetime in quantum ensembles, points to potential applications of optical information processing and optical quantum network with ions, and also an exciting perspective for remote atmospheric coherent optical information storage at room temperature.

As shown in Fig. 1, a femtosecond laser at 800 nm excites the quantum coherence $\rho_{AX}$. Under the perturbation theory in the weak excitation limit, we have $\rho_{AX}(t) \propto iE_1^p \delta t[\rho_{XX}(0) - \rho_{AA}(0)]e^{-\Gamma t}$, where $E_1^p$ is the peak amplitude of the 800 nm laser field, $\delta t$ is the full width at half maximum of the laser field, $\Gamma$ is the dephasing rate, and $\rho_{ii}(0)$ is the initial population at the level *i*. Such strong coherence provides the photon retention and keeps emitting photons near 800 nm with the coherence time up-to picoseconds, with its intensity being approximately described as $I_s(t) \propto \left|E_1^p \delta t[\rho_{XX}(0) - \rho_{AA}(0)]\right|^2 e^{-2\Gamma t}$. Another laser at 1580 nm with the peak amplitude

$E_2^p$ and the same pulse duration is applied at a delay $\tau$, and is used to read-out the re-emitted photon, which triggers a two-photon absorption (TPA) between levels *A* and *B*. The TPA rate can be estimated as $|E_1^p E_2^p \delta t^2|^2 [\rho_{XX}(0) - \rho_{AA}(0)]^2 [\rho_{BB}(0) - \rho_{AA}(0)]^2 e^{-2\Gamma\tau}$. The ultraviolet (UV) radiation from level *B* to *X* is then generated and detected. Since the initial excitation (at 800 nm) prepares very large coherence $\rho_{AX}$ in $N_2^+$ for the photon retention with a relatively long time $\sim 1/\Gamma$ after the 800 nm laser is gone, the re-emitted photons from the coherence can be read-out by a delayed 1580 nm laser through the TPA process within the time window defined by $\sim 1/\Gamma$. In the following, we will first demonstrate the experimental results to validate our hypothesis of photon retention and then support the experimental data with more accurate numerical simulations to further explore the effect.

With the basic concept of our proposal in mind, we designed our experiment in the nitrogen gas. A commercial Ti:sapphire laser (Legend Elite-Duo, Coherent, Inc.) was used to pump an optical parametric amplifier (OPA), which delivered femtosecond laser pulses tunable from 1200 nm to 2400 nm. The remaining fundamental radiation centered at 800 nm was used as a pump beam to promote $N_2^+$ generation. The polarization directions of two laser beams were parallel to each other, and their relative delay was controlled by a motorized translation stage. Two beams were combined using a dichroic mirror and collinearly focused into 5-mbar nitrogen gas using a plano-convex lens (*f* = 20 cm). The peak intensity at the focus was $\sim 2\times 10^{14}$ W/cm$^2$ for the 800-nm, 40-fs laser pulse and $\sim 1\times 10^{13}$ W/cm$^2$ for the 1580-nm, 60-fs laser pulse, where two laser fields were assumed to have Gaussian shapes in both space and time. The UV radiation exiting from the gas chamber was collected and focused onto a slit of a spectrometer (Shamrock 303i, Andor) for spectral analysis. A set of appropriate filters were placed before the spectrometer to block the residual laser radiation.

Figure 2(a) shows the energy diagram for generating the UV radiations through the resonant interaction of two laser fields with $N_2^+$. The 800 nm laser ($\omega_1$) was used for the ionization of $N_2$ molecules and photo-excitation of $N_2^+$ from *X* to *A* state. Afterwards, $N_2^+$ ions were excited to *B* state by absorbing a photon near 800 nm and a photon near 1580 nm ($\omega_2$). The cascaded resonant excitation gave rise to a strong radiation at 329.3 nm. Figure 2(b) shows typical spectra of UV radiations induced by

800 nm and 1580 nm lasers for different time delays: $\tau = 0$ and $\tau = 1$ ps. At $\tau = 0$, a strong radiation at 329.3 nm wavelength was superimposed on a broad spectrum produced by the non-resonant four-wave mixing (FWM). The narrow-bandwidth radiation at 329.3 nm is ascribed to P branch of $B(v=4) \rightarrow X(v=2)$, whereas the side lobe near 328 nm is ascribed to R branch of the same electronic transition. They were generated by resonant FWM of two lasers in $N_2^+$. We observed another narrow-bandwidth radiation at 330.8 nm wavelength, which corresponds to $B(v=2) \rightarrow X(v=0)$ transition. Although the broadband FWM spectrum was also able to cover this transition, the 330.8 nm signal was much weaker than the 329.3 nm radiation. At $\tau = 1$ ps, both the non-resonant FWM signal and 330.8 nm radiation disappeared due to the temporal separation of two pulses. Nevertheless, the 329.3 nm radiation measured at two delays remained almost unchanged.

Figure 2(c) compares the evolution of the non-resonant FWM at 321.0 nm and the resonant FWM at 329.3 nm with the time delay. Unlike the non-resonant FWM which required the temporal overlap of two laser fields, the 329.3 nm signal reached its maximum around the zero delay and almost remained unchanged over the next several picoseconds except for some oscillations probably caused by molecular rotations. When the dynamics was measured over a longer timescale, as illustrated in Fig. 2(d), the 329.3 nm signal exhibited an exponential decay with a time constant of ~26 ps. Generation of the 329.3 nm radiation after the temporal separation of two laser pulses indicates that TPA from $A(v=4)$ to $B(v=4)$ was triggered by the re-emitted photons and the delayed-injected 1580 nm laser.

We further studied the UV radiation driven by the 800 nm laser field and the delay-injected laser with other wavelengths. When the laser wavelength was tuned to 1910 nm, two-photon resonance from $A(v=4)$ to $B(v=4)$ was no longer satisfied. In this case, the narrow-bandwidth radiation at 329.3 nm was submerged by the non-resonant FWM signal [see inset of Fig. 3(a)]. Similar to other spectral components of FWM, the radiation at 329.3 nm was only observed within the temporal overlap of two laser fields, as illustrated in Fig. 3(a). The comparison between Fig. 2(c) and Fig. 3(a) clearly demonstrates that the cascaded resonant condition in the three-level $N_2^+$ system must be fulfilled to generate the strong 329.3 nm radiation after the temporal separation of two laser fields, which agrees with our predication in Fig. 1. When the wavelength of the delayed laser field was tuned to 2070 nm, we could realize one-photon resonant excitation from $X(v=0)$ to $A(v=2)$ and two-photon resonant excitation from $A(v=2)$ to $B(v=2)$ [see Fig. 2(a)]. Although the cascaded resonant condition similar to the 329.3 nm radiation was fulfilled, the 330.8 nm radiation was only observed around zero delay, as shown in Fig. 3(b). This difference for different transitions was beyond our expectation, and hence detailed simulation was necessary to gain a better fundamental understanding.

We performed numerical simulations using Maxwell-Bloch equations. We considered a one-dimensional pencil-like medium with $N_2^+$ modeled by a three-level diagram (i.e., $X(v=2)$, $A(v=4)$, $B(v=4)$) shown in Fig. 2(a). The two-photon excitation of the $A \leftrightarrow B$ transition was treated as two excitation processes via two dipole-allowed transitions with a detuned intermediate level $I$ [35]. Using the rotating-wave approximation, the Hamiltonian in the interaction picture reads [1]

$$H = \hbar\Delta|I\rangle\langle I| + (-\wp_{AX}E_1|A\rangle\langle X| - \wp_{IA}E_1|I\rangle\langle A|$$
$$-\wp_{BI}E_2|B\rangle\langle I| - \wp_{BX}E_s|B\rangle\langle X| + \text{H.c.}), \qquad (1)$$

where $\wp_{ij}$ is the dipole moment between the $i \leftrightarrow j$ transition, $E_{1(2)}$ is the slowly varying envelope amplitude of the 800 (1580) nm electric field, and $\Delta = \omega_{IA} - \omega_1$ is the detuning between $A \leftrightarrow I$ transition and the 800 nm laser field. Assuming $\Delta$ is much larger than any decoherence rate and thus $\rho_{IX}$ is negligible, one can use steady-state values $\rho_{IA} = [\wp_{IA}E_1(\rho_{AA} - \rho_{II}) + (\wp_{BI}E_2)^*\rho_{BA}]/(\hbar\Delta)$ and $\rho_{BI} = [\wp_{BI}E_2(\rho_{BB} - \rho_{II}) + (\wp_{IA}E_1)^*\rho_{BA}]/(\hbar\Delta)$. Hence, the entire light-matter interaction process including two-photon excitation can be described effectively in the density-matrix equations [35]:

$$\dot\rho_{AX} = -\Gamma_{AX}\rho_{AX} + i\frac{\wp_{AX}E_1}{\hbar}(\rho_{XX} - \rho_{AA}) - i\frac{\wp_{BX}E_s}{\hbar}\rho_{AB}, \qquad (2)$$

$$\dot\rho_{BA} = \left(-\Gamma_{BA} + i\frac{|\wp_{BI}E_2|^2 - |\wp_{IA}E_1|^2}{\hbar^2\Delta}\right)\rho_{BA}$$
$$+i\frac{\wp_{BI}E_2\wp_{IA}E_1}{\hbar^2\Delta}(\rho_{AA} - \rho_{BB}) - i\frac{(\wp_{AX}E_1)^*}{\hbar}\rho_{BX} + i\frac{\wp_{BX}E_s}{\hbar}\rho_{XA}, \qquad (3)$$

$$\dot{\rho}_{BX} = -\Gamma_{BX}\rho_{BX} + i\frac{\wp_{BX}E_s}{\hbar}(\rho_{XX} - \rho_{BB}) - i\frac{\wp_{AX}E_1}{\hbar}\rho_{BA}, \quad (4)$$

$$\dot{\rho}_{BB} = -\gamma_B\rho_{BB} + \left(i\frac{\wp_{BI}E_2\wp_{IA}E_1}{\hbar^2\Delta}\rho_{AB} + i\frac{\wp_{BX}E_s}{\hbar}\rho_{XB} + H.c.\right), \quad (5)$$

$$\dot{\rho}_{AA} = -\gamma_A\rho_{AA} + \left(-i\frac{\wp_{BI}E_2\wp_{IA}E_1}{\hbar^2\Delta}\rho_{AB} + i\frac{\wp_{AX}E_1}{\hbar}\rho_{XA} + H.c.\right), \quad (6)$$

$$\rho_{XX} + \rho_{AA} + \rho_{BB} = 1, \quad (7)$$

where dephasing rates are $\Gamma_{AX} = \frac{1}{2}\gamma_A + \gamma_{col}$, $\Gamma_{BA} = \frac{1}{2}(\gamma_A + \gamma_B) + \gamma_{col}$, $\Gamma_{BX} = \frac{1}{2}\gamma_B + \gamma_{col}$, with $\gamma_{col}$ being the collisional dephasing rate, and $\gamma_{A(B)}$ is the spontaneous decay rate from level $A(B)$. The Maxwell equations for field amplitudes read

$$\left(\frac{\partial}{\partial z} + \frac{1}{c}\frac{\partial}{\partial t}\right)\left(\frac{\wp_{BX}E_s}{\hbar}\right) = i\eta_{BX}\rho_{BX}, \quad (8)$$

$$\left(\frac{\partial}{\partial z} + \frac{1}{c}\frac{\partial}{\partial t}\right)\left(\frac{\wp_{AX}E_1}{\hbar}\right) = i\eta_{AX}\rho_{AX}, \quad (9)$$

where $\eta_{iX} = 3n_a\lambda_{iX}^2\gamma_i/8\pi$, $\lambda_{iX}$ is the transition wavelength, and $n_a$ is the $N_2^+$ density. We simulate the FWM signal $E_s$ as well as the $E_1$ field followed by a delayed $E_2$ laser field by solving the Maxwell-Bloch equations (2)-(9). In simulations, we set parameters: the medium length as 0.15 mm, $\gamma \sim 0.01$ ns$^{-1}$, $\gamma_{col} \sim 1$ ns$^{-1}$, $\wp \sim 10^{-30}$ C·m, $\Delta \sim 2\pi * 10^{15}/2\pi$ Hz, and $n_a = 4 \times 10^{16}$ cm$^{-3}$. Peak amplitudes of two original lasers are set as $E_1^p \sim 3 \times 10^{10}$ V/m and $E_2^p \sim 0.7 \times 10^{10}$ V/m, respectively. Both two laser pulses have a Gaussian intensity envelope with the pulse duration of 50 fs. The length chosen here is smaller for the simplicity of simulations, which shall not hinder the verification of physics. All other parameters are chosen roughly at the same order as those in the experiment.

We plot numerical results for evolutions of the 800 nm laser and $\rho_{BA}$ for the cases of $\tau = 0$ and $\tau = 1$ ps in Fig. 4. Results for different initial population $\rho_{BB}(0)$ on level $B$ are considered. After the original 800 nm laser passes through the medium, the calculated laser field shows a long tail with oscillations. The tail is 7~8 orders of magnitude weaker than the main peak, and the intensity ratio of the main peak and the tail is almost independent of $\rho_{BB}(0)$, as shown in Fig. 4(a). The tail means that $\rho_{AX}$ is generated and continuously releases weak emission around 800 nm, which is consistent with the theory that photon retention happens, i.e., photons are stored in the coherence and re-emitted. Such re-emitted weak photons in the medium can then be read-out when the 1580 nm laser field is injected with the time delay. For the case of $\tau = 1$ ps, the coherence $\rho_{BA}$ between levels $A$ and $B$ arises due to the two-photon excitation process and reaches a larger amplitude as $\rho_{BB}(0)$ increases [see Fig. 4(c)]. Although $\rho_{BA}$ is much smaller at $\tau = 1$ ps than that at $\tau = 0$, it is still possible to generate 329.3 nm radiation through the resonant FWM provided the re-emitted photons near 800 nm keeps interacting with ions.

To give a straightforward comparison with experimental results, we plotted calculated spectra of signal fields and integrals of signal intensities versus delay for different $\rho_{BB}(0)$ values. Signal field spectra for simultaneous pumping from two fields in Fig. 5(a) show a Fano line shape, while the signal field at $\tau = 1$ ps evolves in a large time scale resulting in a narrow spectral line in Fig. 5(b). In contrast to the case of $\tau = 0$, the 329.3 nm signal at $\tau = 1$ ps grows dramatically as $\rho_{BB}(0)$ increases. This dependence is also shown in curves of the total signal versus delay. As shown in Fig. 5(c), a larger $\rho_{BB}(0)$ results in a smaller decrease of the intensity versus the delay. For the case of $\rho_{BB}(0) = 0$, the 329.3 nm signal at $\tau = 1$ ps is about 10 orders of magnitude weaker than that at $\tau = 0$. Such a weak signal is impossible to be measured experimentally. Furthermore, after the initially quick drop, the decay rate of the signal becomes smaller, which is consistent with the experimental result in Fig. 2(c). The simulation results show that $B$-state population is of vital importance for reading-out the re-emitted photon. The high initial population on the $B$(v=4) state actually triggers larger coherence between levels $B$ and $A$ during the two-photon resonance procedure, allowing us to experimentally observe the photon retention effect through the coherence-enhanced radiation at 329.3 nm.

Actually, the $B$(v=4) level can be efficiently populated through three-photon near-resonant excitation shown in Fig. 2(a), which has been demonstrated in previous simulations [30,32]. To confirm this fact experimentally, we injected a femtosecond seed pulse around 330 nm after the 800 nm laser. Although the seed pulse can cover all transitions $B(v=2,3,4) \rightarrow X(v=0,1,2)$ with $\Delta v = 2$, it is only amplified at 329.3 nm, as shown in Fig. 3(c), indicating that the $B(v=4)$ level has a higher population than the $X(v=2)$ level. For the three-level system composed of $X(v=2)$, $A(v=4)$ and $B(v=4)$, the high population probability on $B(v=4)$ level makes the 329.3 nm radiation easily be obtained with two temporal separated laser fields. In comparison, the absorption at 330.8 nm means higher population on $X(v=0)$ level than the $B(v=2)$ level. This is because the $B(v=2)$ level cannot be efficiently populated through the three-photon absorption from the $X(v=0)$ level with the 800 nm laser field or the strong-field ionization. Low population on the $B(v=2)$ level makes the 330.8 nm radiation cannot be efficiently generated after the separation of two lasers, which is consistent with the experimental result in Fig. 3(b). Our numerical simulation results convincingly prove the proposed photon retention and show excellent qualitative agreement with experimental measurements. Further improvement in simulations is possible if we include more realistic beam profiles of lasers, inhomogeneous density distribution of $N_2^+$, and the dynamic ionization process, which, however, increases the numerical difficulty.

To conclude, we have demonstrated for the first time the photon retention and optical readout in the quantum optical platform made out of nitrogen ions. Photons near 800 nm are stored and re-emitted from the quantum coherence in $N_2^+$, which are re-emitted and then read-out by applying a delayed 1580 nm laser to induce the UV radiation. Moreover, large population on the $B(v=4)$ level is produced through the three-photon excitation, which largely enhances the read-out signal. Our work reveals that photons re-emitted by strong excited coherence can be used to perform nonlinear process in ionized nitrogen molecules, which triggers a broad interest for potential applications involving those ions such as remote sensing [29], quantum computation [36], as well as quantum simulations [37], and holds possibility for serving as an important tool targeting towards biological imaging [38] and surface spectroscopy [39,40]. Our pioneering experiment performed in molecular nitrogen with tunnel ionization also opens an important route towards future potential implementation of light storage in remote atmospheric regime.

**Acknowledgements**

This work is supported by National Natural Science Foundation of China (11822410, 12034013, 11734009, 11974245); National Key R&D Program of China (2017YFA0303701, 2019YFA0705000); Shanghai Municipal Science and Technology Major Project (2019SHZDZX01); Program of Shanghai Academic Research Leader (20XD1424200); Natural Science Foundation of Shanghai (19ZR1475700); Strategic Priority Research Program of Chinese Academy of Sciences (XDB16030300); Key Research Program of Frontier Sciences of Chinese Academy of Sciences (QYZDJ-SSW-SLH010); Youth Innovation Promotion Association of CAS (2018284); NSF (ECCS-1509268, CMMI-1826078) and AFOSR (FA9550-20-1-0366, FA9550-20-1-0367). This work is also partially supported by the Fundamental Research Funds for the Central Universities. L. Y. acknowledges support from the Program for Professor of Special Appointment (Eastern Scholar) at Shanghai Institutions of Higher Learning. X. C. also acknowledges the support from Shandong Quancheng Scholarship (00242019024).

**Author contributions**

Jinping Yao, Luqi Yuan, Xianfeng Chen, and Ya Cheng discussed and conceived the idea. Jinping Yao designed the experiment. Jinming Chen and Jinping Yao performed the experiments. Luojia Wang and Luqi Yuan performed the simulations. Jinping Yao and Luqi Yuan analyzed the data. The manuscript was prepared by Jinping Yao, Luojia Wang, Vladislav V. Yakovlev, Luqi Yuan, Xianfeng Chen, and Ya Cheng, and was discussed among all authors.

**Captions of figures:**

Fig. 1. (Color online) (a) Schematic for achieving the photon retention with $N_2^+$ ions and (b) the corresponding energy-level schemes. The 800 nm laser excites the electronic coherence between levels $X$ and $A$, which keeps emitting photons near 800 nm. A delayed 1580 nm laser is used to readout the re-emitted photon via TPA, which then generates the UV radiation.

Fig. 2. (Color online) (a) Energy diagram of the resonant excitation in $N_2^+$ for generating UV radiations at 329.3 nm and 330.8 nm. (b) Spectra of UV radiations induced by the 800 nm and 1580 nm lasers in the cases of $\tau = 0$ and $\tau = 1$ ps. (c) The evolution of resonant FWM at 329.3 nm and non-resonant FWM at 321.0 nm with the time delay. (d) The temporal evolution of the 329.3 nm radiation measured versus the time delay over the 100 ps timescale.

Fig. 3. (Color online) (a) The 329.3 nm radiation induced by 800 nm and 1910 nm lasers versus the time delay. The spectrum at zero delay is shown in inset. (b) The 330.8 nm radiation induced by 800 nm and 2070 nm lasers versus the time delay. The spectrum around zero delay is shown in inset. (c) The spectrum obtained by injecting the 800 nm laser and a 330 nm seed with the 0.5 ps delay into 80 mbar nitrogen gas. The radiation and absorption peaks ascribed to $B(v=2,3,4)\rightarrow X(v=0,1,2)$ with $\Delta v = 2$ are indicated with black arrows, and the corresponding vibrational states are shown in the brackets. Red dotted line indicates the seed spectrum.

Fig. 4. (Color online) (a) Intensity of the calculated output 800 nm ($E_1$) pulse with initial population $\rho_{BB}(0) = 0, 0.1, 0.2, 0.4$. The 1580 nm ($E_2$) pulse injected at the 1 ps delay is indicated by the orange filled curve. Temporal evolutions of $|\rho_{BA}|$ for the cases of (b) $\tau = 0$ and (c) $\tau = 1$ ps with $\rho_{BB}(0) = 0, 0.1, 0.2, 0.4$.

Fig. 5. (Color online) Calculated spectra of the signal in the cases of (a) $\tau = 0$ and (b) $\tau = 1$ ps with $\rho_{BB}(0) = 0, 0.1, 0.2, 0.4$. (c) Integral of the signal intensity versus $\tau$ for different $\rho_{BB}(0)$.

Fig. 1

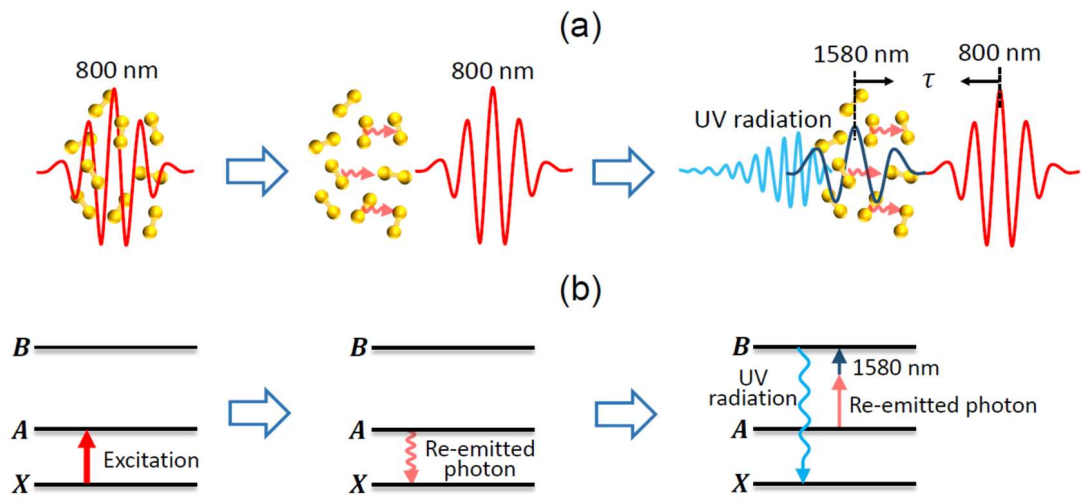

Fig. 2

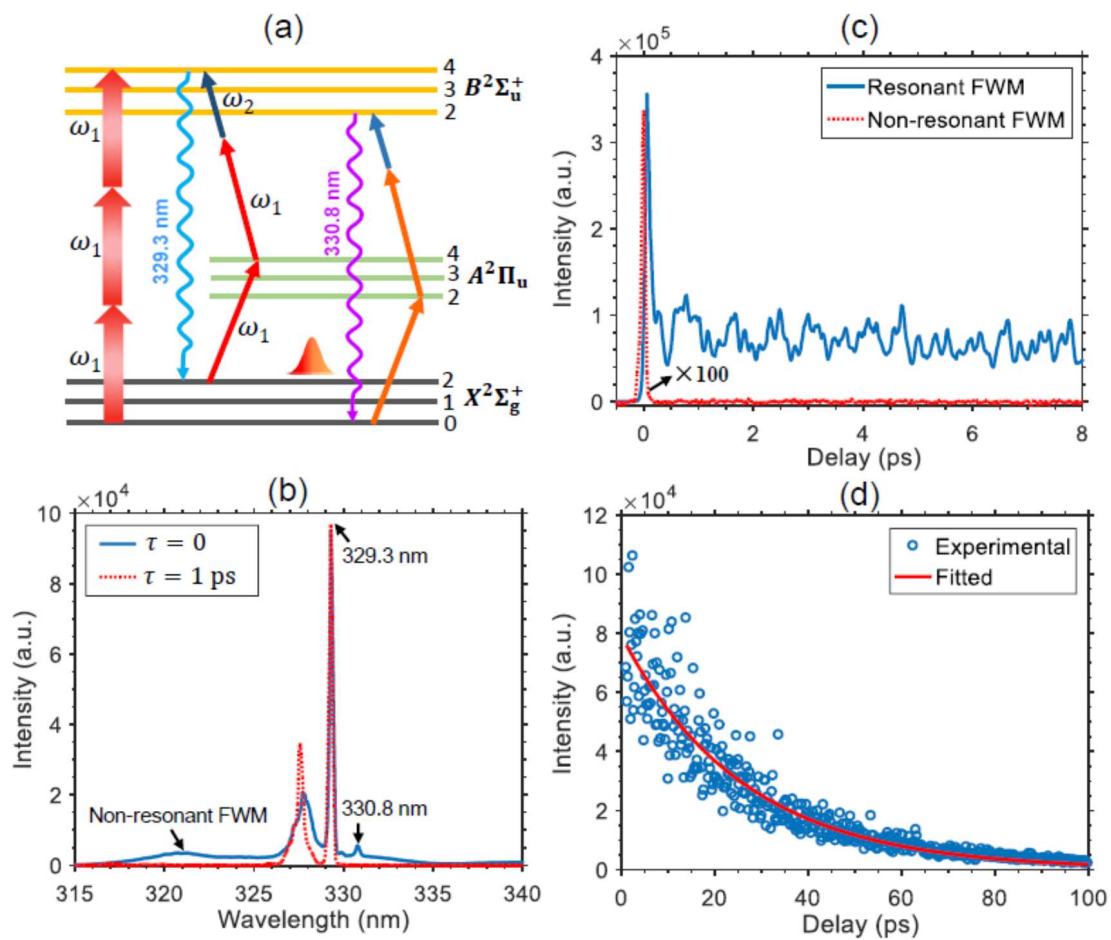

Fig. 3

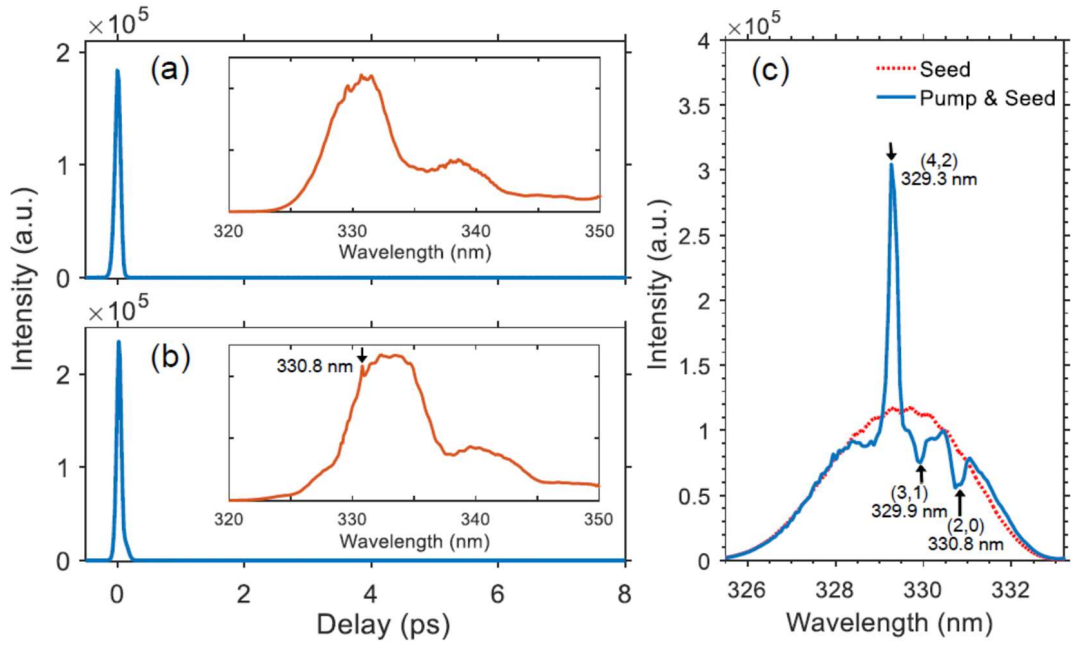

Fig. 4

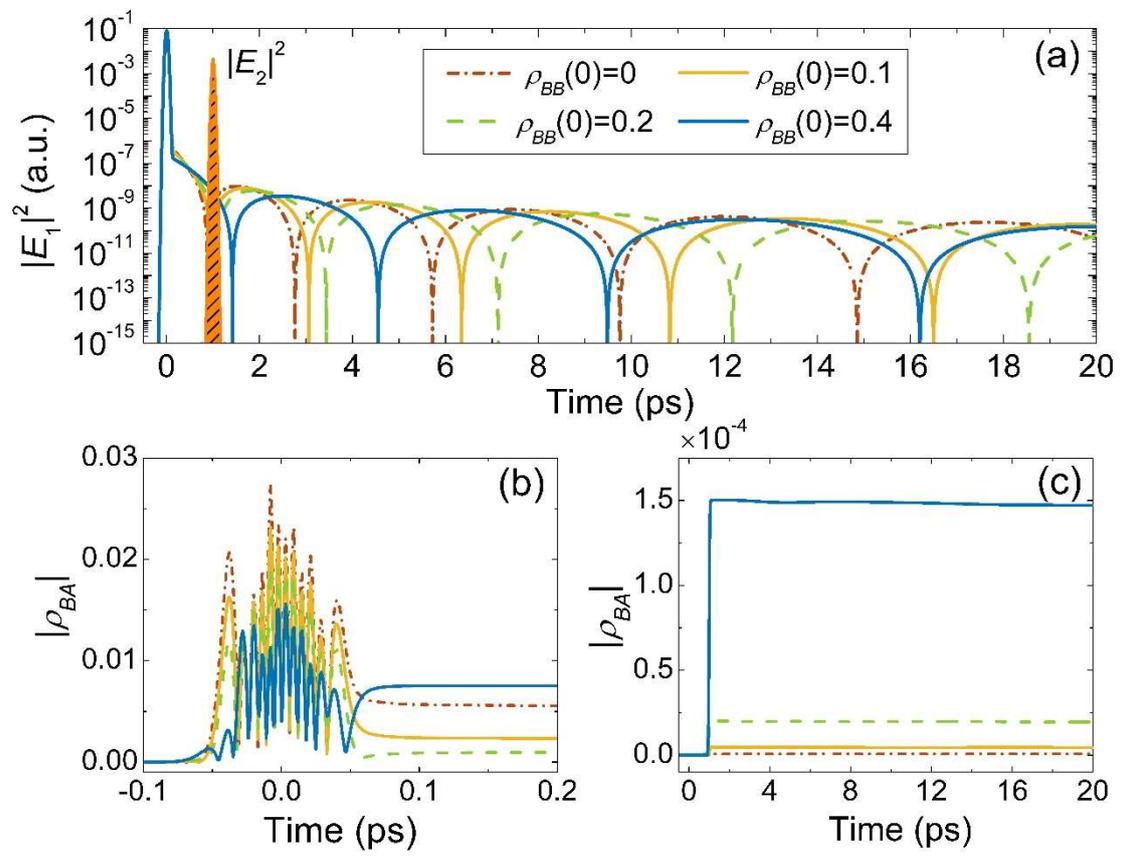

Fig. 5

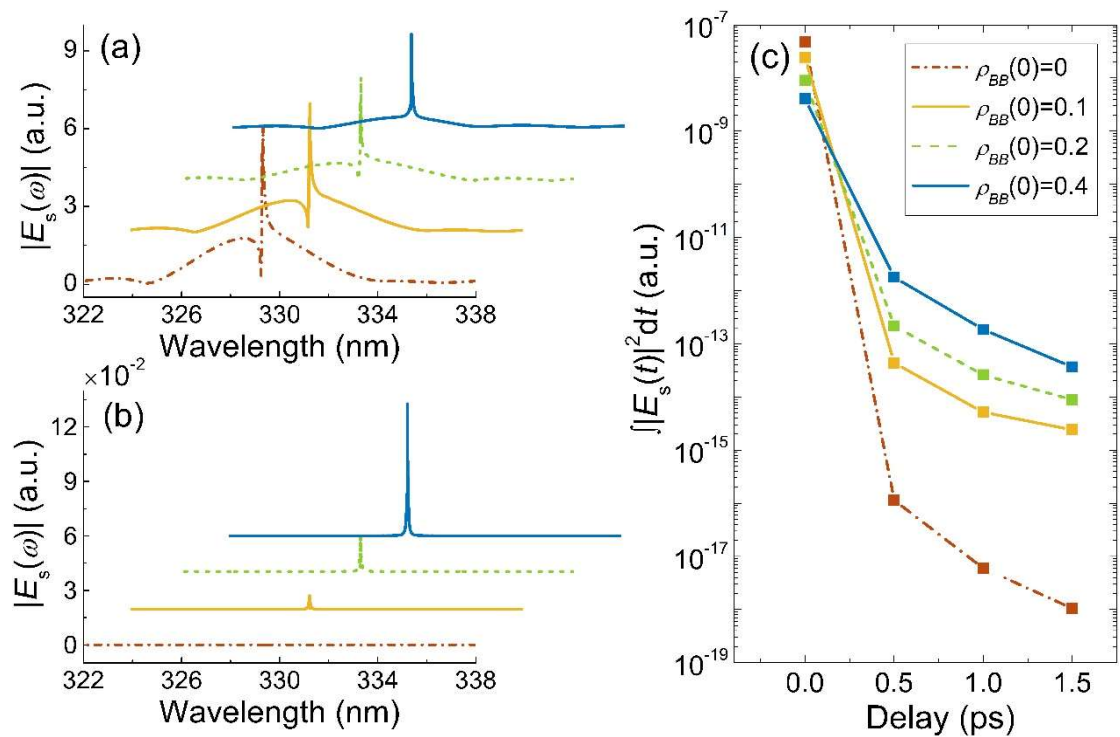